\documentclass[preprint]{aastex}
\usepackage{gensymb}
\def\degree{${}^{\circ}$}
\slugcomment{Submitted to ApJL}
\shortauthors{Zheng et al.}
\begin{document}

\title{Compound eruptions of twin flux ropes in a solar active region}
\author{Ruisheng Zheng$^{1,2}$, Liang Zhang$^{1}$, Bing Wang$^{1}$, Xiangliang Kong$^{1}$, Hongqiang Song$^{1}$, Zhao Wu$^{1}$, Shiwei Feng$^{1}$, Huadong Chen$^{2}$, and Yao Chen$^{1}$}
\affil{$^{1}$Shandong Key Laboratory of Optical Astronomy and Solar-Terrestrial Environment, School of Space Science and Physics, Institute of Space Sciences, Shandong University, Weihai, Shandong, 264209, China; ruishengzheng@sdu.edu.cn\\
$^{2}$CAS Key Laboratory of Solar Activity, National Astronomical Observatories, Beijing 100012, China\\}

\begin{abstract}
Compound eruptions represent that multiple closely spaced magnetic structures erupt consecutively within a short interval, and then lead to a single flare and a single CME. However, it is still subtle for the links between multiple eruptions and the associated single flare or/and single CME. In this Letter, we report the compound eruptions of twin close flux ropes (FR1 and FR2) within a few minutes that resulted in a flare with a single soft X-ray peak and a CME with two cores. The successive groups of expanding loops and double peaks of intensity flux in AIA cool wavelengths indicate two episodes of internal magnetic reconnections during the compound eruptions. Following the eruption of FR2, the erupting FR1 was accelerated, and then the expanding loops overlying FR2 were deflected. Moreover, the eruption of FR2 likely involved the external magnetic reconnection between the bottom of the overlying stretching field lines and the rebounding loops that were previously pushed by the eruption of FR1, which was evidenced by a pair of groups of newly-formed loops. All results suggest that the compound eruptions involved both internal and external magnetic reconnections, and two erupting structures of twin FRs interacted at the initial stage. We propose that two episodes of internal magnetic reconnections were likely united within a few minutes to form the continuous impulsive phase of the single peaked flare, and two separated cores of the CME was possibly because that the latter core was too slow to merge with the former one.
\end{abstract}


\section{Introduction}
Magnetic flux ropes (FRs) are characterized by groups of coherently twisted magnetic field lines that collectively wind around the central axis. In EUV or soft X-ray passbands, FRs can appear as hot channels or sigmoids representing hot plasma tracing the helical magnetic field lines (Zhang et al. 2012; Liu 2020). FRs are generally considered as core structures of the most magnificent coronal eruptions, solar flares and coronal mass ejections (CMEs). Hence, the eruptions of FRs release a tremendous amount of magnetic energy previously stored in the solar atmosphere, and can be one main driver for adverse space weather near the Earth (Chen 2017; Cheng et al. 2017; Gibson 2018; Wang \& Liu 2019; Liu 2020).

The eruptions of FRs play an important role to understand the physical mechanisms of solar eruptions. In the standard eruption model (Carmichael 1964; Sturrock 1966; Hirayama 1974; Kopp \& Pneuman 1976; Aulanier et al. 2012, 2013; Janvier et al. 2013, 2015), a rising FR stretches the overlying magnetic field lines; as a result, a current sheet forms beneath the rising FR between the stretching anti-parallel field lines; then, successive magnetic reconnections occur in the current sheet, and finally cause the growing flare loops in the source and help the FR quickly evolve into a CME propagating into the outer corona. During the eruptions, the free magnetic energy accumulated in the corona is quickly transformed into bulk motions, plasma heating, and kinetic energy of nonthermal particles (Priest \& Forbes 2002; Shibata \& Magara 2011).

It is familiar for an eruption of a single FR, but more and more eruptions are frequently found to involve multiple FRs (T{\"o}r{\"o}k et al. 2011; Liu et al. 2012; Li \& Zhang 2013; Kliem et al. 2014). If multiple FRs share a same polarity inversion line (PIL) under the common closed flux system, they can be stacked vertically as a double-decker FR (Liu et al. 2012; Cheng et al. 2014; Zheng et al. 2017; Hou et al. 2018; Mitra et al. 2020), and can also braided about each other (Awasthi et al. 2018). When multiple FRs lie above different PILs in a complex active region (AR) or different ARs, they can disrupt respective closed flux systems and produce sympathetic eruptions (T{\"o}r{\"o}k et al. 2011; Chintzoglou et al. 2015; Song et al. 2020).

If multiple FRs resided closely and successively erupted within a short period, the multiple eruptions will be compound and be accompanied with a single flare and a single CME (Dhakal et al. 2018). It remains elusive for the links between the compound eruptions and the associated single flare and CME. In this Letter, we present the compound eruptions of twin FRs within a few minutes in a same AR (SOL2015-12-23T00:15:34L319C115), and study the relationships between the compound eruptions and the associated single flare and CME.

\section{Observations}
Twin FRs (FR1 and FR2) situated closely in NOAA AR 12473 ($\sim$S21E65), and erupted successively within a few minutes on 2015 December 23. Hence, the successive eruptions were compound, and were linked to a single peaked flare of M4.7 class, a CME of $\sim$520 km s$^{-1}$, and a type II radio burst. The main observations are from the Atmospheric Imaging Assembly (AIA; Lemen et al. 2012) onboard the Solar Dynamics Observatory (SDO; Pesnell et al. 2012). The AIA instrument's 10 EUV and UV wavelengths involve a wide range of temperatures. The AIA images cover the full solar disk and up to 0.5 $R_\odot$ above the limb, with a pixel resolution of 0$\farcs$6 and a cadence of 12 s. We also used full-disk magnetograms from the Helioseismic and Magnetic Imager (HMI; Scherrer et al. 2012) also onboard SDO, with a cadence of 45 s and pixel scale of 0$\farcs$6 to examine the magnetic field configuration of the source region. The evolution of the CME in the high corona was captured by Large Angle and Spectrometric Coronagraph (LASCO; Brueckner et al. 1995) onboard the {\it Solar and Heliospheric Observatory (SOHO)} spacecraft. In addition, the associated type II radio burst was detected in the frequency range of 25-180 MHz by the metric spectrometer from Learmonth (LEAR; Kennewell \& Steward 2003).

We also investigated the emission properties of the FRs with the differential emission measure (DEM) method by employing the sparse inversion code (Cheung et al. 2015; Su et al. 2018). In DEM method, the EM maps at different temperature ranges are obtained by a set of AIA images in six channels (i.e., 94, 131, 171, 193, 211, and 335~{\AA}).


\section{Results}
\subsection{Compound Eruptions of twin flux ropes}
The compound eruptions of twin FRs are shown in AIA 131~{\AA} images and EM maps in different temperature ranges (Figure 1 and Animation 1). $\sim$3 minutes before the eruption (Figure 1(a)), FR1 (the red arrow) appeared as a bright structure that obstructed most part of FR2, and only the top portion of FR2 was identified (the pink arrow). FR1 first erupted at $\sim$00:25 UT, and the eruption of FR1 made the hidden FR2 be totally visible as a twisted loop (the red and pink arrows in Figure 1(b)). In a few minutes, FR2 also erupted following the ejecting FR1 (Figure 1(c)). It is noticeable that a bundle of loops (L1; blue arrows) was pushed northward due to the evacuation of FR1 (Figure 1(a)-(c)). The erupting FR1 and FR2 were very clear in the EM maps at the temperature range of 8-12.5 MK (red and pink arrows in Figure 1(d)-(f)). In addition, FR2 had a clear hotter kernel in the EM maps at the temperature range of 12.5-25 MK (pink arrows in Figure 1(g)-(h)).

The compound eruptions of twin FRs over the limb are well displayed in composite running-ratio-difference images in AIA 171 (red), 193 (green), and 131 (blue)~{\AA} (Figure 2(a)-(e) and Animation 2). $\sim$5 minutes before the eruption, FR1 (the red arrow) lay lowly in the AR, and FR2 (the pink arrow) suspended highly over the AR (Figure 2(a)). At the commencement of the eruption, FR1 quickly rose and obscured FR2 (Figure 2(b)). The profiles of FR1 and FR2 are overlaid in the red box, and the closeup of twin FRs is much clear in the black box at the left-down corner (Figure 2(a)-(b)). After the successive eruptions, two groups of expanding loops successively appeared over the limb. The first group of expanding loops (L2; the white arrow) was intimately associated with the erupting FR1 (the red arrow), while FR2 (the pink arrow) was still at a lower height (Figure 2(c)). The sudden appearance of the second group of expanding loops (L3; the green arrow) was closely associated with the erupting FR2 (the pink arrow), while the erupting FR1 and the expanding L2 nearly left the field of view (FOV) of AIA (Figure 2(d)). Interestingly, the main propagation direction of the erupting FR2 was changed clockwise by $\sim$20$\degree$ (S1-S2; the blue and black lines starting from diamonds), which possibly indicated that the erupting FR2 experienced a deflection. Note that a pair of groups of higher loops (L4; the black arrow) and lower loops (L5; the orange arrow) formed in the meanwhile (Figure 2(e)). Since the formation, L3 kept expanding, while L4 nearly stayed still (green and black arrows in Figure 2(e)).

In addition, the compound eruptions were related to a type II radio burst as an indicator of the formation of a coronal shock. In the radio spectrum of LEAR (Figure 2(f)), the type II radio burst appeared as two clear strong stripes that correspond to the fundamental (F) and harmonic (H) branches, with the onset at $\sim$00:36 UT (the dashed line). On the other hand, following the compound eruptions and the loop expansions, a single CME with a speed of $\sim$520 km s$^{-1}$ was detected by LASCO ({\url{https://cdaw.gsfc.nasa.gov/CME\_list/UNIVERSAL/2015\_12/univ2015\_12.html}}). The CME front (the white arrow) and the bright core (the red arrow) first emerged beneath the streamer (the cyan arrow) into the FOV of LASCO/C2 at about one hour later than the eruption of FR1 (Figure 2(g)). Interestingly, one more hour later ($\sim$02:36 UT), another weaker core appeared in the FOV of LACSO/C2 and slowly displayed a helical structure in three hours (pink arrows in Figure 2(f)-(g)).

\subsection{External Magnetic Reconnection}
Next, we focus on the formation of the pair of L4 and L5 (Figure 3 and Animation 3). As showed in Figure 1, L1 was pushed northward during the eruption of FR1. After the evacuation of FR1, L1 (the blue arrow) immediately bounced back and got close to a ray-like structure (the yellow arrow) lying along the eruption direction (Figure 3(a)). During the return of L1, some jets (green arrows) emanated from the eruption core (Figure 3(b)). As soon as the contact with the ray-like structure, L1 disappeared, and the lower L5 formed (the orange arrow in Figure 3(c)), consistent with the appearance of the pair of L4 and L5 in Figure 2(b). It is likely that external magnetic reconnection occurred between the rebounding L1 and the ray-like structure and produced L4 and L5. The rebounding L1 and the jet was clear in the EM maps at the temperature range of 0.5-4 MK (the blue and green arrows in Figure 3(d)-(e)), and the ray-like structure appeared as a current sheet in the EM map at the temperature range of 12.5-25 MK (the yellow arrow in Figure 3(f)). On the other hand, the compound eruptions resulted in an M4.7 solar flare that involved with two separate groups of flare loops that shared a common southern end (Figure 3(g)-(i)). The first (red arrows) and second (pink arrows) groups of flare loops respectively related to the eruptions of FR1 and FR2, and the second group appeared a few minutes later but was much stronger. The current sheet (yellow arrows) obviously extended outwards from the top of the second group of flare loops.

For the flare region (the box in Figure 3(g)), we check the intensity evolution in AIA EUV wavelengths (Figure 4(a)). It is clear that the intensity curves for AIA 211, 335, and 94~{\AA} have a single peak at $\sim$00:48 UT (the dotted line), and the curves for AIA 304, 171, 193, and 131~{\AA} have two peaks at $\sim$00:37 (the dashed line) and $\sim$00:48 UT. The single peak in hotter passbands was consistent with the single flare peak (the red vertical line) that was shown in the GOES soft X-ray flux (Figure 4(b)). Note that the flare peak of $\sim$00:40 UT was a few minutes earlier than the peak (the black dashed line) in AIA 94~{\AA} (Figure 4(a)-(b)).

The kinematic evolutions for the expanding L2-L3 and erupting FRs along the initial eruption direction and the deflection direction (S1-S2 in Figure 2) are shown in the time-distance plots of AIA running-ratio-difference images (Figure 4(c)-(f)). Along the initial eruption direction of S1, the erupting FR1 and the overlying L2 experienced a clear acceleration from $\sim$250 km s$^{-1}$ to $\sim$480 km s$^{-1}$, and the erupting FR2 and the overlying L3 had a constant speed of $\sim$250 km s$^{-1}$. The accelerations of FR1 and L2 began at $\sim$00:31 and $\sim$00:33 UT (the red and white vertical lines), respectively. Along the deflection direction of S2, L2 also speeded up from $\sim$235 km s$^{-1}$ to $\sim$400 km s$^{-1}$, and L3 similarly expanded with a constant speed of $\sim$270 km s$^{-1}$. It is clear that FR1 and FR2 appeared a few minutes later than them along S1. FR1 appeared after its acceleration onset (the red dashed line in Figure 4(d)), and therefore only showed the later high speed of $\sim$400 km s$^{-1}$. Absorbingly, FR2 displayed two moving branches with speeds of $\sim$240 and $\sim$270 km s$^{-1}$, while it nearly vanished along S1. The kinematic evolution for the rebounding L1 along the return direction (S3; the white line starting from the white diamond in Figure 2(c)) is shown in the time-distance plots of AIA normalised-intensity images (Figure 4(g)-(h)). L1 consisted of two branches of loops (blue and cyan arrows), and we mainly analyse the main branch (blue arrows). L1 was first deflected from the eruption region at a speed of $\sim$50 km s$^{-1}$. Then, L1 began to bounce back at a quicker speed of $\sim$130 km s$^{-1}$ at $\sim$00:31 UT (the blue vertical line), close to the onset of the FR2 eruption. Finally, L5 appeared at $\sim$00:35 UT (the orange vertical line) as soon as the disappearance of L1, and immediately swayed with a deceleration from $\sim$80 km s$^{-1}$ to $\sim$60 km s$^{-1}$.

\section{Conclusions and Discussion}
Combining with the AIA images, EM maps, and LASCO coronagraphs, it is clear that the successive eruptions of twin FRs in AR 12473 was compound within a few minutes on 2015 December 23. The first eruption of FR1 pushed away a group of nearby loops (L1), and the rope-like structures with hot and dense kernels were confirmed in AIA hot wavelengths and EM maps at the temperature ranges of 8-12.5 and 12.5-25 MK (Figure 1). Consequently, the compound eruptions successively blew two groups of loops (L2 and L3) up ahead of twin erupting FRs, and was closely related to a type II radio burst, and finally evolved into a CME with two cores (Figure 2). During the eruptions, two groups of flare loops were produced in the source region. It is intriguing that a pair of groups of higher and lower loops (L4 and L5) simultaneously formed as soon as the interaction between the rebound L1 and the current sheet, and the current sheet rooted at the the top of the stronger group of flare loops at a temperature range of 12.5-25 MK (Figure 3).

The compound eruptions of FR1 and FR2 naturally caused two close groups of flare loops, which is consistent with double peaks of intensity flux in AIA 304, 171, 193, and 131~{\AA} (the dashed and dotted line in Figure 4(a)). However, there is only one single peak of intensity flux in AIA 211, 335, and 94~{\AA} that was some minutes later than the single peak in GOES soft X-ray flux (the dashed line in Figure (b)). Considering the cooler and hotter components of the characteristics temperature in AIA 131 and 193~{\AA}, we can understand that one common peak ($\sim$00:48 UT) exists for all AIA EUV passbands, and there is one more peak ($\sim$00:31 UT) for AIA cool wavelengths. Therefore, the first peak in cool wavelengths possibly indicates that magnetic reconnection between the anti-parallel field lines stretched by the erupting FR1 occurred very low, and the latter peak for all AIA EUV wavelengths shows that magnetic reconnection between the anti-parallel field lines stretched by the erupting FR2 happened in higher locations, which was confirmed by the current sheet atop the higher second group of flare loops (Figure 3). It is possible that the first peak in hot wavelengths indeed existed, but was not as discernible as that in cool wavelengths. On the other hand, the onsets of two eruptions (the red and pink solid lines) are both between the start and the peak of the related flare (Figure 4(b)). Hence, due to the short interval of compound eruptions, it is likely that two episodes of energy release were united within a few minutes to form a continuous impulsive phase of the single peaked flare.

The acceleration start of the erupting FR1 and the overlying L2 is almost coincident with the onset of the FR2 eruption (dashed lines in Figure 4(c)-(d)). However, it is peculiar that the eruption speed of FR2 with a stronger energy release was smaller than that of FR1 with a weaker energy release. How the slowly-moving FR2 accelerated the fast-moving FR1 (Figure 4)? It is possible that the erupting structure of FR2 was restricted by the previously-erupted structure of FR1 and then accelerated the erupting structure of FR1 during its slowly-rising phase, due to their intimate temporal and spacial relationship (Figure 2(a)-(c)), and the previously-erupted FR1 was still nearby the overlying L3 of FR2 at the beginning of the eruption of FR2 (the red and green triangles in Figure 4(c)). Furthermore, the expanding direction of the overlying L3 was changed clockwise by $\sim$20$\degree$ (Figure 2(e)), which indicates that the erupting structure of FR2 was deflected after the confinement from the erupting structure of FR1. As a result, the deflected FR2 showed two branches along S2 but disappeared along S1 in the time-distance plots (Figure 4(d) and (f)), and the contours of deflected FR2 are overlaid in Figure 2(d)-(e). Hence, we suggest that thee interaction between two erupting structures resulted in the slowness of FR2 and the acceleration of FR1 at the initial stage.

In addition, the type II radio burst occurred a few minutes later than the onset of the FR2 eruption (the dashed line in Figure 2(f)). It is likely that the eruption of FR2 contributed for the acceleration of the L2 as the leading front of the recorded CME, and then a type II radio burst was triggered at a height of 1.29 $\sim$$R_\odot$ (the triangle in Figure 2(d)). At this height of the AR outskirts, it is possible that the front with a speed of $\sim$480 km s$^{-1}$ (the triangle in Figure 4(c)) triggered a type II radio burst in the ambient medium with a lower characteristic (magnetosonic or Alfv{\'e}n) speed ($\sim$300 km s$^{-1}$) (Gopalswamy et al. 2001; 2009). Interestingly, L1 began to deflect simultaneously with the start of the FR1 rise (the blue and red solid lines in Figure 4(d)-(f)), but kept rebounding during the eruption of FR2 (the blue and red dashed lines in Figure 4(d)-(f)). It is possible that FR1 erupted from a lower height to pushed the nearby L1, but the eruption of FR2 from a higher location did not affect the rebounding motion of L1. The deduced original heights of two FRs are consistent with the configuration of two groups of flare loops (Figure 3). Intriguingly, the pair of L4 and L5 formed as soon as the interaction between the rebounding L1 and the current sheet. The formation onset of L5 (and L4) was only one minute later than the appearance start of the expanding L3 (the green and orange solid lines in Figure 4(c) and 4(h)), which possibly indicates that external magnetic reconnection occurred between the L1 and the field lines wrapping the current sheet at the bottom of the stretching L3.

Based on the observational results and the discussions, we propose a possible scenario for the compound eruptions in both top (Figure 5(a)-(d)) and edge (Figure 5(e)-(h)) views superimposed on the HMI magnetogram. Before the eruption (Figure 5(a) and 5(e)), the low-lying FR1 (red ropes) and the high-lying FR2 (pink ropes) resided closely over two different PILs, and the L1 (cyan lines) stood at the AR east edge. Incidentally, the truth for the positive polarity connecting the south end of L1 was confirmed by tracking its movement toward disk center in the following days. After the eruption of FR1 (Figure 5(b) and 5(f)), L1 was pushed away, and the overlying field lines (L2; blue lines) were stretched, and the internal magnetic reconnection occurred in a lower current sheet (the red shadow) at the bottom of the stretched L2. For the eruption of FR2 (Figure 5(c)-(g)), the internal magnetic reconnection occurred in a higher current sheet (the pink shadow) at the bottom of the stretched overlying field lines (L3; yellow lines), and the the external magnetic reconnection (red pentangles) happened between the rebounding L1 and the bottom field lines of the stretched L3. Finally (Figure 5(d) and 5(h)), the internal magnetic reconnections produced two groups of flare loops (red and pink arcades), and the external magnetic reconnection resulted in the formation of L4 (green lines) and L5 (orange lines). Note that the erupting structure of FR2 initially moved toward previously-erupted structure of FR1 (the green arrow), and then deflected away (Figure 5(f)-(g)).

On the other hand, the internal magnetic reconnections of the compound eruptions created two CMEs in the scenario (red and pink circles and expanding blue and yellow lines in Figure 5). It is consistent with that FR1 and FR2 located above different PILs, different from the double-decker FR. Though twin FRs successively erupted within a few minutes, two cores did not merge together, possibly because the second core was much slow than the first one. Therefore, it was only detected for one CME with one leading front and two cores (Figure 2(g)-(i)). The first bright core and the fast front correspond to the erupting FR1 and the overlying expanding L2 after the acceleration, and the slow second helical core represents the slow erupting FR2. Due to the compound eruptions propagated at a similar direction in the same streamer, the invisibility of the second front ahead of the second core was possibly resulted from the local depletion of the first CME. In addition, the external magnetic reconnection between L1 and L3 can also weaken the front (Figure 5(c)).

In summary, twin FRs resided closely over two different PILs, and their successive eruptions were compound within a short interval (7 minutes). The primary aim of this Letter is to understand the relationship between the compound eruptions of twin FRs (a low-lying FR1 and a high-lying FR2) and the associated single CME and single flare. All results suggest that two erupting structures of twin FRs interacted at the initial stage, and the compound eruptions involved both internal and external magnetic reconnections. We propose that two episodes of internal magnetic reconnections were likely united within a few minutes to form the continuous impulsive phase of the single peaked flare, and two separated cores of the CME was possibly because that the latter core was too slow to merge with the former one. The details of compound eruptions can complement the understanding of the initiation and evolution of multiple eruptive structures. More and better observations are necessary to verify the suggestions.

\acknowledgments
The authors thank the anonymous referee for constructive comments and Leping Li for the helpful discussion. R. Zheng would like to appreciate his wife (Xinchen Li) and his daughter (Shuyu Zheng) for the long time support. {\it SDO} is a mission of NASA's Living With a Star Program. The authors thank the teams of {\it SDO}, SOHO, and LEAR for providing the data. This work is supported by grants NSFC 11790303 and 12073016, and the open topic of the Key Laboratory of Solar Activities of Chinese Academy Sciences (KLSA202108).

\clearpage

\begin{figure}
\epsscale{0.95} \plotone{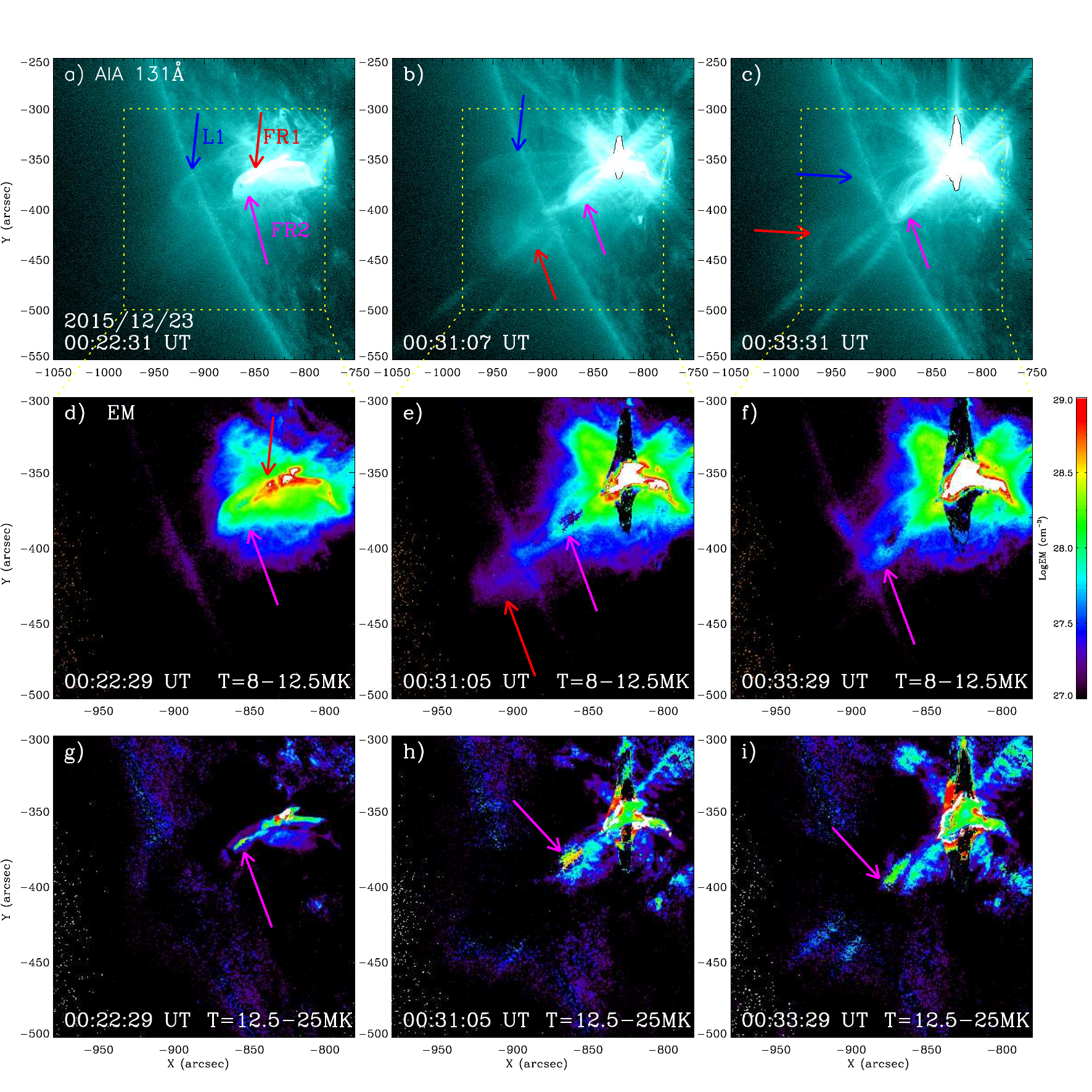}
\caption{Eruptions of FR1 (the red arrow) and FR2 (the pink arrow) in AIA 131~{\AA} (top panels) and in EM maps at temperature range of 8-12.5 MK (middle panels) and 12.5-25 MK (bottom panels). The blue arrows indicated L1, and the dashed boxes represent the FOV of panels (d)-(i). (An animation of AIA images and EM maps is available online. The animated sequence runs from 00:20 to 00:35 UT.)
\label{f1}}
\end{figure}

\clearpage

\begin{figure}
\epsscale{1.0}
\plotone{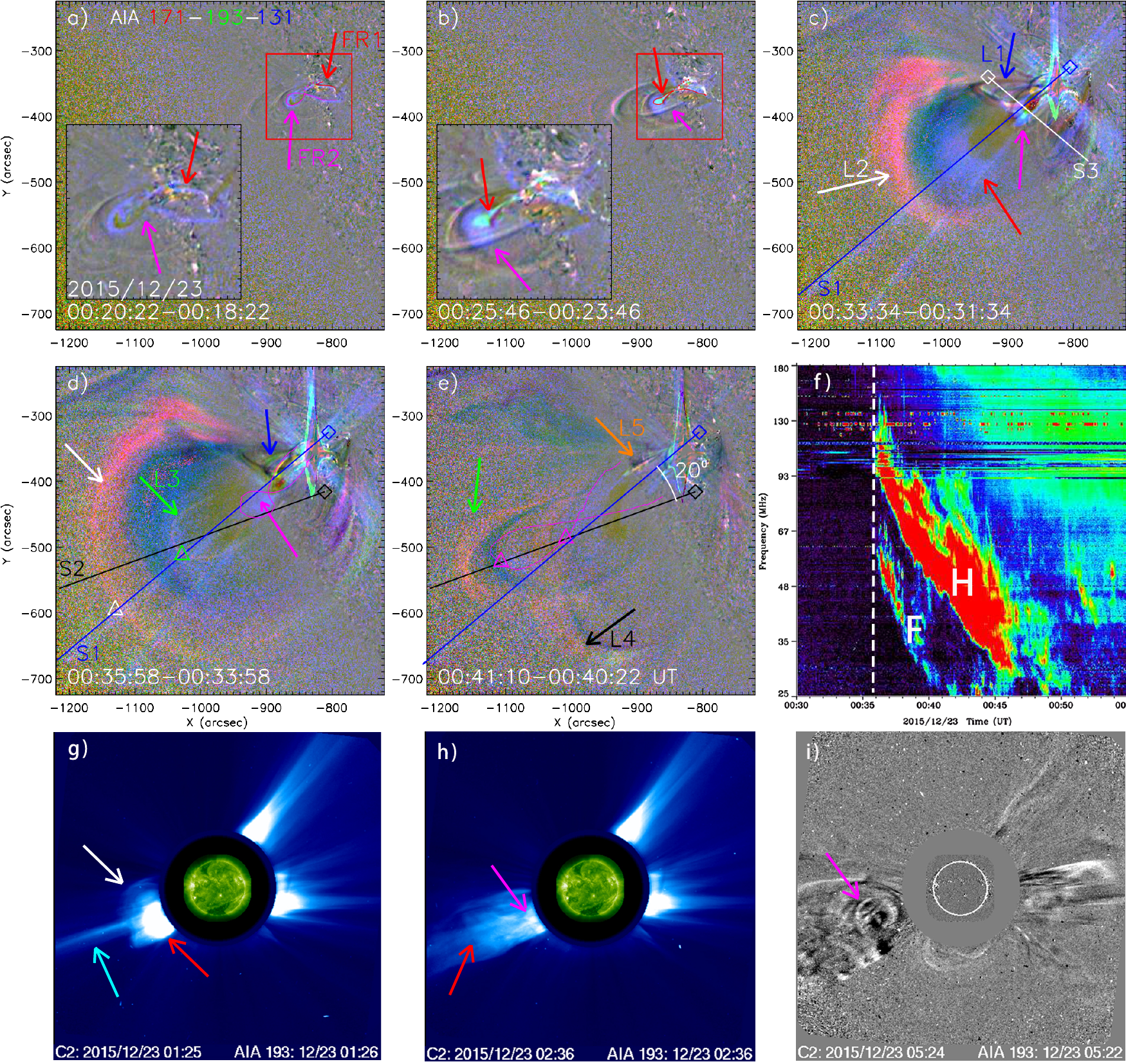}
\caption{((a)-(e)) Eruptions of FR1 (the red arrow) and FR2 (the pink arrow) in composite running-ratio-difference images of AIA 171 (red), 193 (green), and 131 (blue)~{\AA}. In panels (a)-(b), the red and pink curves represent the contours of FR1 and FR2, and the red boxes show the FOV of insets. The associated loops are indicated by the blue, white, green, black, and orange arrows, respectively. The lines starting at diamonds are used to derive the attached speeds in Figure 4. (f) The fundamental (F) and harmonic (H) branches of the type II radio burst in the dynamic spectrum of LEAR (25-180 MHz). The dashed line shows the start of the burst. ((g)-(i)) The CME emanating below the streamer (the cyan arrow) in images of LASCO C2. The white, red, and pink arrows indicate the leading front and two cores, respectively. (An animation of composite running-ratio-difference AIA images is available online. The animated sequence runs from 00:20 to 00:50 UT.)
\label{f2}}
\end{figure}

\clearpage

\begin{figure}
\epsscale{0.95}
\plotone{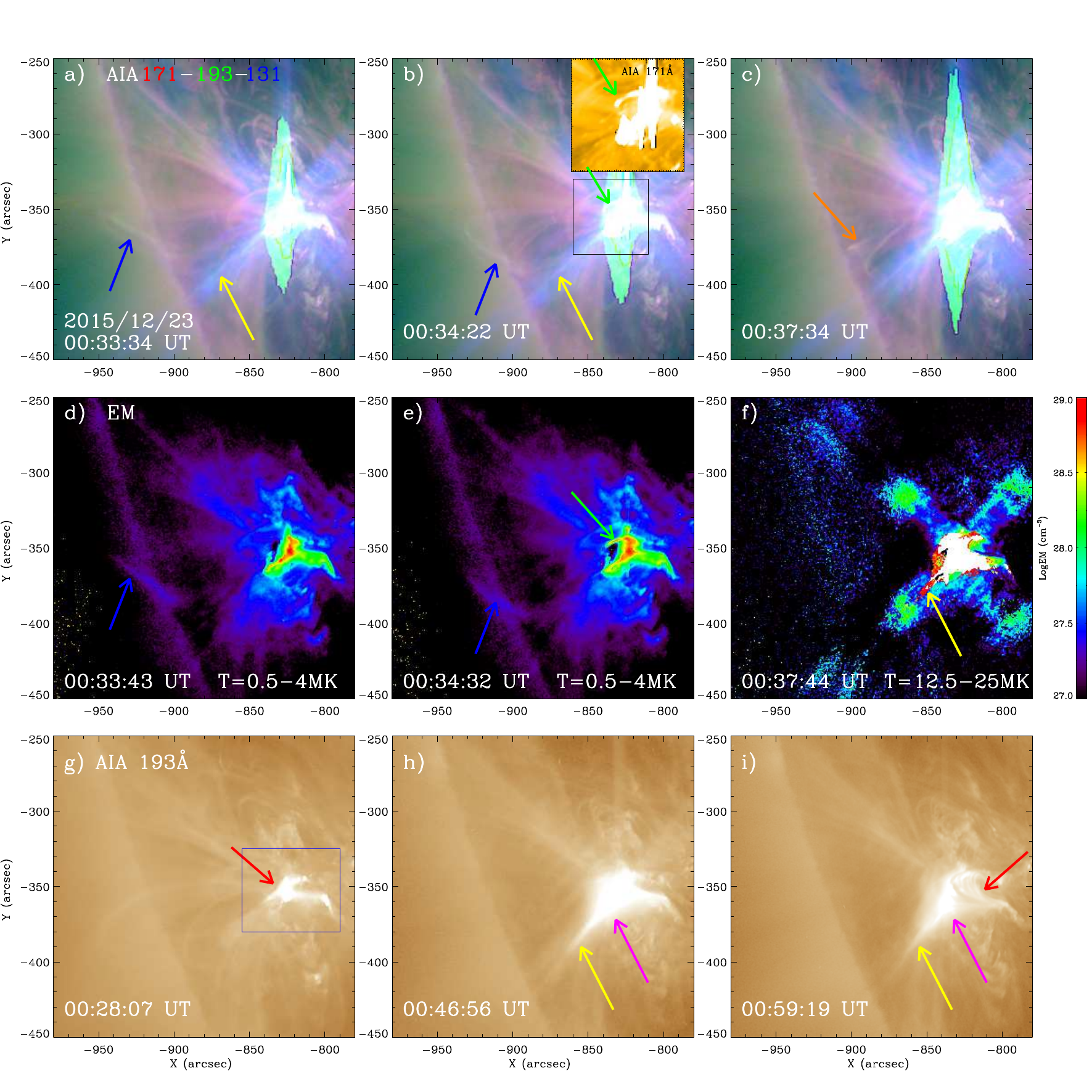}
\caption{Magnetic reconnection between the rebound L1 (blue arrows) and the current sheet (yellow arrows) in composite images of AIA 171 (red), 193 (green), and 131 (blue)~{\AA} (top panels) and in EM maps at temperature range of 8-12.5 MK and 0.5-4 MK (middle panels) and flare loops (red and pink arrows) in AIA 193~{\AA} (bottom panels). The green and orange arrows separately indicate the jet and the newly-formed L5, and the blue box represents the flare region. The black box shows the FOV of the inset in panel (b). (An animation of AIA images and EM maps is available online. The animated sequence runs from 00:30 to 00:40 UT.)
\label{f3}}
\end{figure}

\clearpage

\begin{figure}
\epsscale{0.5}
\plotone{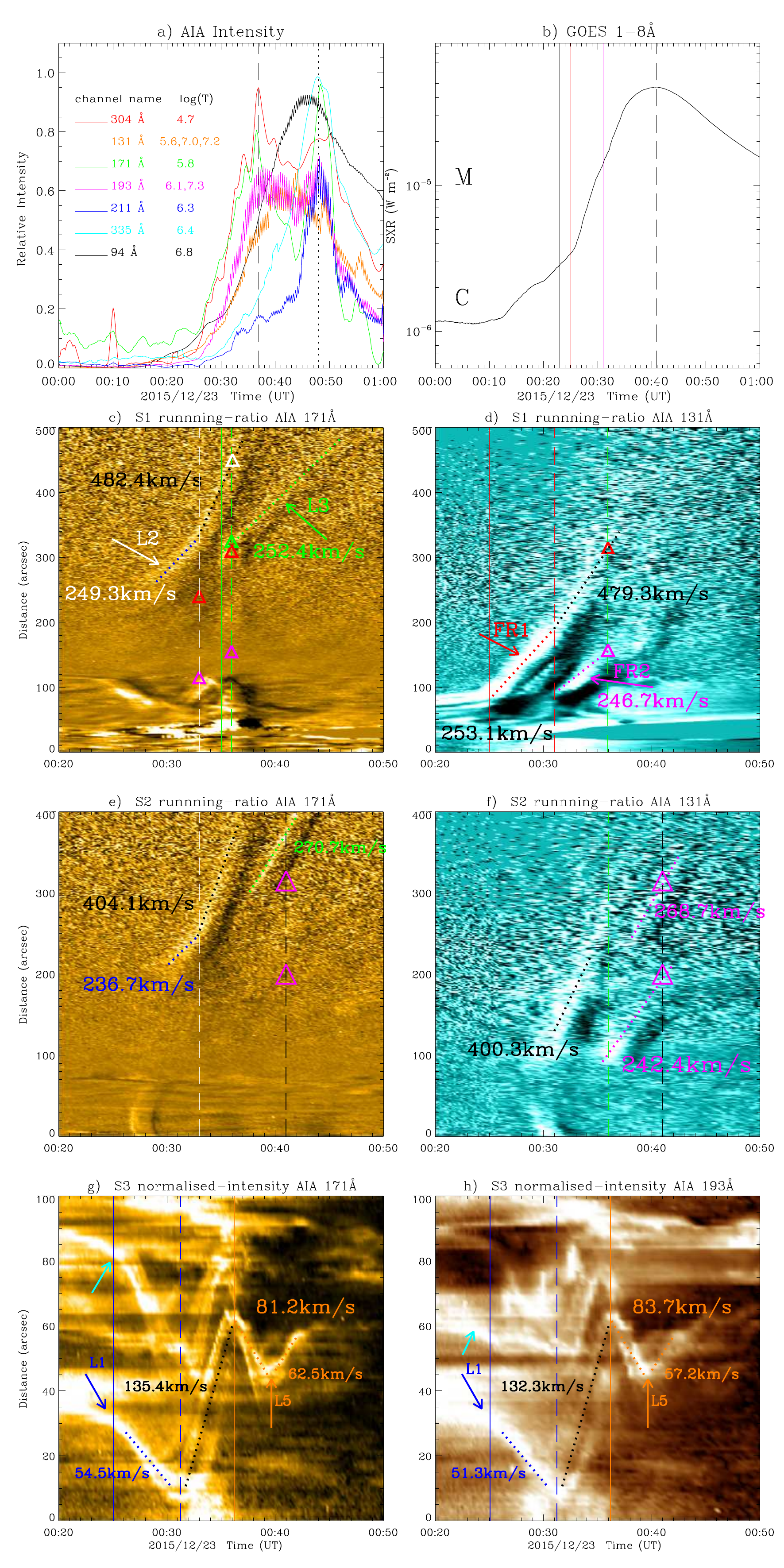}
\caption{(a) The intensity curves in seven AIA wavelengths for the flare region (the box in Figure 3). The dashed and dotted lines indicate two peak times. (b) GOES soft X-ray flux for the flare. The black vertical lines mark the beginning and peak times of the flare, and the red and pink lines show the onsets of twin eruptions. ((c)-(f)) Time-distance plots along S1 and S2 in Figure 2 of running-ratio-difference images in AIA 171 and 131~{\AA}, uncovering the propagations of FR1 (the red arrow), FR2 (the pink arrow), L2 (the white arrow), and L3 (the green arrow). ((g)-(h)) Time-distance plots along S3 in Figure 2 of normalized-intensity-original images in AIA 171 and 193~{\AA}, showing the movements of L1 (cyan and blue arrows) and L5 (the orange arrow). The solid and dashed lines indicate some timing of the associated propagation and movements. The white, green, red, and pink triangles represent the locations of L1, L2, FR1, and FR2. The dotted lines are used to derive the attached speeds. .
\label{f4}}
\end{figure}

\clearpage

\begin{figure}
\epsscale{0.9}
\plotone{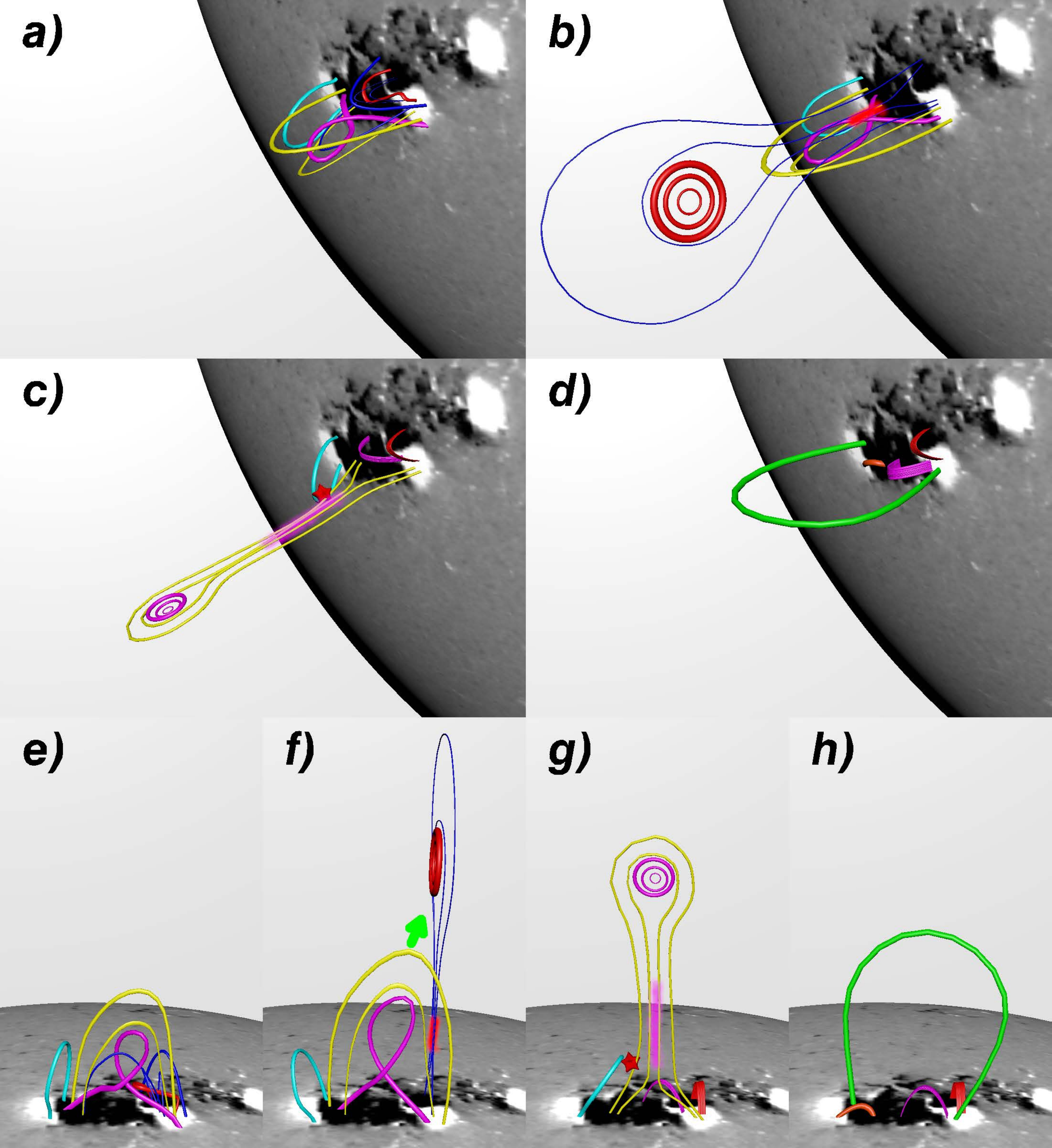}
\caption{Scenario for the compound eruptions of FR1 (red ropes) and FR2 (pink ropes) in both top (panels (a)-(d)) and edge (panels (e)-(h)) views superimposed on the HMI magnetograms. L1-L5 are indicated by the cyan, blue, yellow, green, and orange lines, respectively. The CME cores and flare loops are marked as red and pink circles and arcades. The shadows and pentangles represent the sites of magnetic reconnections.
\label{f5}}
\end{figure}


\begin{thebibliography}{}

\bibitem{aulanier13} Aulanier, G., D{\'e}moulin, P., Schrijver, C. J., et al. 2013, \aap, 549, A66

\bibitem{aulanier12} Aulanier, G., Janvier, M., \& Schmieder, B. 2012, \aap, 543, A110
\bibitem{awasthi18} Awasthi, A. K., Liu, R., Wang, H., Wang, Y., \& Shen, C. 2018, \apj, 857, 124

\bibitem[Brueckner et al. (1995)]{brueckner95} Brueckner, G. E., Howard, R. A., Koomen, M. J., Korendyke, C. M. et al. 1995, \solphys, 162, 357

\bibitem{carmichael64} Carmichael, H. 1964, in the Physics of Solar Flares, Proceedings of the AAS-NASA Symposium, 50, A Process for Flares, ed. Wilmot N. Hess, 451

\bibitem{chen17} Chen, J. 2017, Physics of Plasmas, 24, 090501

\bibitem{cheng14} Cheng, X., Ding, M. D., Zhang, J., et al. 2014, \apj, 789, 93
\bibitem{cheng17} Cheng, X., Guo, Y. \& Ding, M. D. 2017, Science China Earth Sciences, 60, 1383

\bibitem{cheung15} Cheung, M. C. M., Boerner, P., Schrijver, C. J., et al. 2015, \apj, 807, 143

\bibitem{chintzouglou15} Chintzoglou, G., Patsourakos, S., \& Vourlidas, A. 2015, \apj, 809, 34

\bibitem{dhakal18} Dhakal, S. K., Chintzoglou, G., \&Zhang, J. 2018, \apj, 860, 35

\bibitem{Gopalswamy01} Gopalswamy, N., Lara, A., Kaiser, M.L., Bougeret, J.-L. 2001, J. Geophys. Res. 106, 25261.
\bibitem{Gopalswamy09} Gopalswamy, N., Thompson, W. T., Davila, J. M., Kaiser, M. L., Yashiro, S. et al. 2009, \solphys, 259, 227.

\bibitem{hirayama74} Hirayama, T. 1974, \solphys, 34, 323

\bibitem{hou18} Hou, Y., Zhang, J., Li, T., Yang, S., \& Li, X. 2018, \aap, 619, A100

\bibitem{janvier15} Janvier, M., Aulanier, G., \& D{\'e}moulin, P. 2015, \solphys, 290, 3425
\bibitem{janvier13} Janvier, M., Aulanier, G., Pariat, E., \& D{\'e}moulin, P. 2013, \aap, 555, A77

\bibitem{kenne03} Kennewell, J. \& Steward, G. 2003, Solar Radio Spectrograph [SRS] Data Viewer [srsdisplay] (Sydney: IPS Radio and Space Serv.)

\bibitem{kliem14} Kliem, B., T{\"o}r{\"o}k, T., Titov, V. S., et al. 2014, \apj, 792, 107

\bibitem{kopp76} Kopp, R. A., \& Pneuman, G. W. 1976, \solphys, 50, 85

\bibitem{lemen12} Lemen, James R., Title, Alan M., Akin, David J., Boerner, Paul F. et al. 2012, \solphys, 275, 17


\bibitem{Li13} Li, L. \& Zhang, J. 2013, \aap, 552, L11

\bibitem{liu20} Liu, R. 2020,Research in Astronomy and Astrophysics, 20, 165

\bibitem{liu12} Liu, R., Kliem, B., T{\"o}r{\"o}k, T., et al. 2012, \apj, 756, 59

\bibitem{mitra20} Mitra, P. K., Joshi, B., Veronig, A. M., Chandra, R., Dissauer, K., \& Wiegelmann, T. 2020, \apj, 900, 23

\bibitem{pesnell12} Pesnell, W. Dean, Thompson, B. J., Chamberlin, P. C. et al. 2012, \solphys, 275, 3

\bibitem{preiest02} Priest, E. R., \& Forbes, T. G. 2002, A\&A Rev., 10, 313

\bibitem[Scherrer et al. (2012)]{scherrer12} Scherrer, P. H., Schou, J., Bush, R. I., Kosovichev, A. G., et al. 2012, \solphys, 275, 207

\bibitem{Shibata2011} Shibata, K. \& Magara, T. 2011, Living Reviews in Solar Physics, 8, 6

\bibitem{song20} Song, Z., Hou, Y., Zhang, J, \& Wang, P. 2020, \apj, 892, 79
\bibitem{sturrock66} Sturrock, P. A. 1966, Nature, 211, 695

\bibitem{su18} Su, Y., Veronig, A. M., Hannah, I. G., et al. 2018, \apjl, 856, L17

\bibitem{torok11} T{\"o}r{\"o}k, T., Panasenco, O., Titov, V. S., et al. 2011, \apjl, 739, L63

\bibitem{Wang} Wang, H., \& Liu, C. 2019, Frontiers in Astronomy and Space Sciences, 6, 18

\bibitem{zhang12} Zhang, J., Cheng, X. \& Ding, M. D. 2012, Nature Comm., 3, 747

\bibitem{zheng17} Zheng, R., Zhang, Q., Chen, Y., Wang, B., Du, G., Li, C. \& Yang, K. 2017, \apj, 836, 160






\end{thebibliography}
\end{document}